\documentclass[galaxies,article,accept,moreauthors,pdftex,12pt,a4paper]{mdpi}
\lastpage{x}
\doinum{10.3390/------}
\pubvolume{xx}
\pubyear{2015}
\history{Received: xx / Accepted: xx / Published: xx}
\pdfoutput=1

\Title{Creating S0s with major mergers: a 3D view}

\Author{Miguel Querejeta $^{1,}$*, M.~Carmen Eliche-Moral $^{2}$, Trinidad Tapia $^{3}$, Alejandro Borlaff $^{4}$, Glenn van de Ven $^{1}$, Mariya Lyubenova $^{5}$, Marie Martig $^{1}$, Jes\'{u}s Falc\'{o}n-Barroso $^{4}$, Jairo M\'{e}ndez-Abreu~$^{6}$, Jaime Zamorano $^{2}$, and Jes\'{u}s Gallego $^{2}$}

\address{%
$^{1}$ Max Planck Institute for Astronomy, K\"{o}nigstuhl 17, D-69117 Heidelberg, Germany\\
$^{2}$ Departamento de Astrof\'{i}sica y Ciencias de la Atm\'{o}sfera, Universidad Complutense de Madrid, E-28040 Madrid, Spain\\
$^{3}$ Instituto de Astronom\'{i}a, Universidad Nacional Aut\'{o}noma de M\'{e}xico, BC-22800 Ensenada, Mexico\\
$^{4}$ Instituto de Astrof\'{i}sica de Canarias, C/ V\'{i}a L\'{a}ctea, E-38200 La Laguna, Tenerife, Spain\\
$^{5}$ Kapteyn Astronomical Institute, University of Groningen, Postbus 800, NL-9700 Groningen, The Netherlands\\
$^{6}$ School of Physics and Astronomy, University St Andrews, North Haugh, St Andrews, KY16 9SS, UK}

\corres{email: querejeta@mpia-hd.mpg.de; telephone: +49 6221 528 335; fax: +49 6221 528 246.}

\abstract{A number of simulators have argued that major mergers can sometimes preserve discs \citep[e.g.][]{2005ApJ...622L...9S}, but the possibility that they could explain the emergence of lenticular galaxies (S0s) has been generally neglected. In fact, observations of S0s reveal a strong structural coupling between their bulges and discs, which seems difficult to reconcile with the idea that they come from major mergers. However, in \citet{2015A&A...573A..78Q} we have used \textit{N}-body simulations of binary mergers to show that, under favourable conditions, discs are first destroyed but soon regrow out of the leftover debris, matching observational photometric scaling relations \citep[e.g.][]{2010MNRAS.405.1089L}.
Additionally, in \citet{2015A&A...579L...2Q} we have shown how the merger scenario agrees with the recent discovery that S0s and most spirals are not compatible in an angular momentum--concentration plane. This important result from CALIFA constitutes a serious objection to the idea that spirals transform into S0s mainly by fading (e.g.~via ram-pressure stripping, as that would not explain the observed simultaneous change in $\lambda_\mathrm{Re}$ and concentration), but our simulations of major mergers do explain that mismatch. From such a 3D comparison we conclude that mergers must be a relevant process in the build-up of the current population of S0s.}



\begin{document}


\section{Introduction}

Ever since \citet{1936rene.book.....H} placed them between ellipticals and spirals in his tuning-fork diagram, lenticular galaxies (or S0s) have been considered as a transition class and have not deserved too much attention until recently. However, lenticulars represent the majority of early-type galaxies in the local Universe \citep{1991rc3..book.....D}, and as such they constitute one of the most important end-products of galaxy evolution.

Recent efforts have shown that S0s actually form a whole sequence, parallel to that of spirals, with bulge-to-total ratios and rotational support levels which span a large range of values, in a similar fashion to the change from early- to late-type spiral galaxies \citep{2010MNRAS.405.1089L,2011MNRAS.416.1680C,2012ApJS..198....2K}.
In this context, the question arises as to whether this parallelism in properties is the consequence of an underlying genetic connection; in other words, are the different types of lenticulars (S0a, S0b, S0c), the end-products of a morphological transformation starting from the corresponding spiral galaxies (Sa, Sb, Sc)?

This possibility has been successfully explored in the cluster regime as a consequence of ram pressure stripping (as a spiral galaxy falls into a cluster, the pressure exerted by the intra-cluster medium can expel most of the gas from the galaxy, suppressing star formation and transforming it into a passive lenticular; \citep{2005AJ....130...65C,2006A&A...458..101A,2008AJ....136.1623C, 
 2015MNRAS.447.1506M}). This would be a so-called \textit{fading} mechanism, and it would indirectly contribute to explaining the observational morphology-density relation \citep[e.g.][]{1980ApJ...236..351D}.

However, not all lenticulars live in clusters; today we know that probably at least half of the S0 population is located in the less dense environment of groups \citep{2009ApJ...692..298W, 2011MNRAS.415.1783B}, where ram pressure is negligible, and mergers become increasingly relevant \citep{2014ApJ...782...53M,2014AdSpR..53..950M}. The aim of the talk given in the context of the European Week of Astronomy (EWASS) in Tenerife, June 2015, was to show how \textit{major} mergers can transform spiral galaxies into lenticulars, in agreement with the tight photometric relations observed for S0s \citep{2010MNRAS.405.1089L}, and explaining the  change in concentration and angular momentum found by CALIFA (van de Ven et al. in preparation). The presentation was mostly based on two recently published papers \citep{2015A&A...573A..78Q,2015A&A...579L...2Q}.

\section{Creating lenticulars with major mergers: photometry}

One of the strongest arguments given so far against a possible merger origin of lenticular galaxies are the photometric scaling relations observed in real S0s, particularly in terms of bulge-to-disc coupling. \citet{2010MNRAS.405.1089L} have found tight scaling relations between the size (and brightness) of bulges and discs in lenticular galaxies, measured through photometric structural decomposition of near-infrared images (\textit{K}-band), and interpret them as the result of secular evolution, arguing that mergers would destroy that structural coupling. However, by performing analogous photometric decompositions on the remnants of simulations of major galaxy mergers, we have found that under favourable conditions discs can be rebuilt during the last stages of the encounter, and the resulting disc galaxies \textit{do obey} the photometric scaling relations of S0s. 

The possiblity of disc survival during major mergers of spiral galaxies had already been suggested by some simulations (e.g.~\citep{2005ApJ...622L...9S, 2009ApJ...691.1168H}). However, ours is the first study in which the remnants are consistently translated into the phase space of observations (through mock images which take the appropriate, variable mass-to-light ratio into account, and performing the relevant photometric decompositions). We point the interested reader to \citet{2015A&A...573A..78Q} for details on the procedure and for exhaustive results and discussion, and we only attempt to provide a short summary next.

\subsection{Simulations of major mergers and mock images}

The public GalMer database\footnote{GalMer project: http://galmer.obspm.fr} provides a comprehensive suite of \textit{N}-body numerical simulations of galaxy mergers,  sampling a wide range of mass ratios, morphology, and orbital characteristics. With a spatial resolution of 0.28\,kpc, $\sim 10^5$ particles per galaxy, and a powerful TreeSPH code, the simulations carefully consider the effects of gas and star formation, with timescales spanning a total of 3--4\,Gyr (up to $\sim$1\,Gyr of relaxation time after the merger is completed); for more details, see \citep{2010A&A...518A..61C}.

\begin{figure*}[t]
\begin{center}
\includegraphics[trim=20 150 10 0,clip,height=0.57\textwidth]{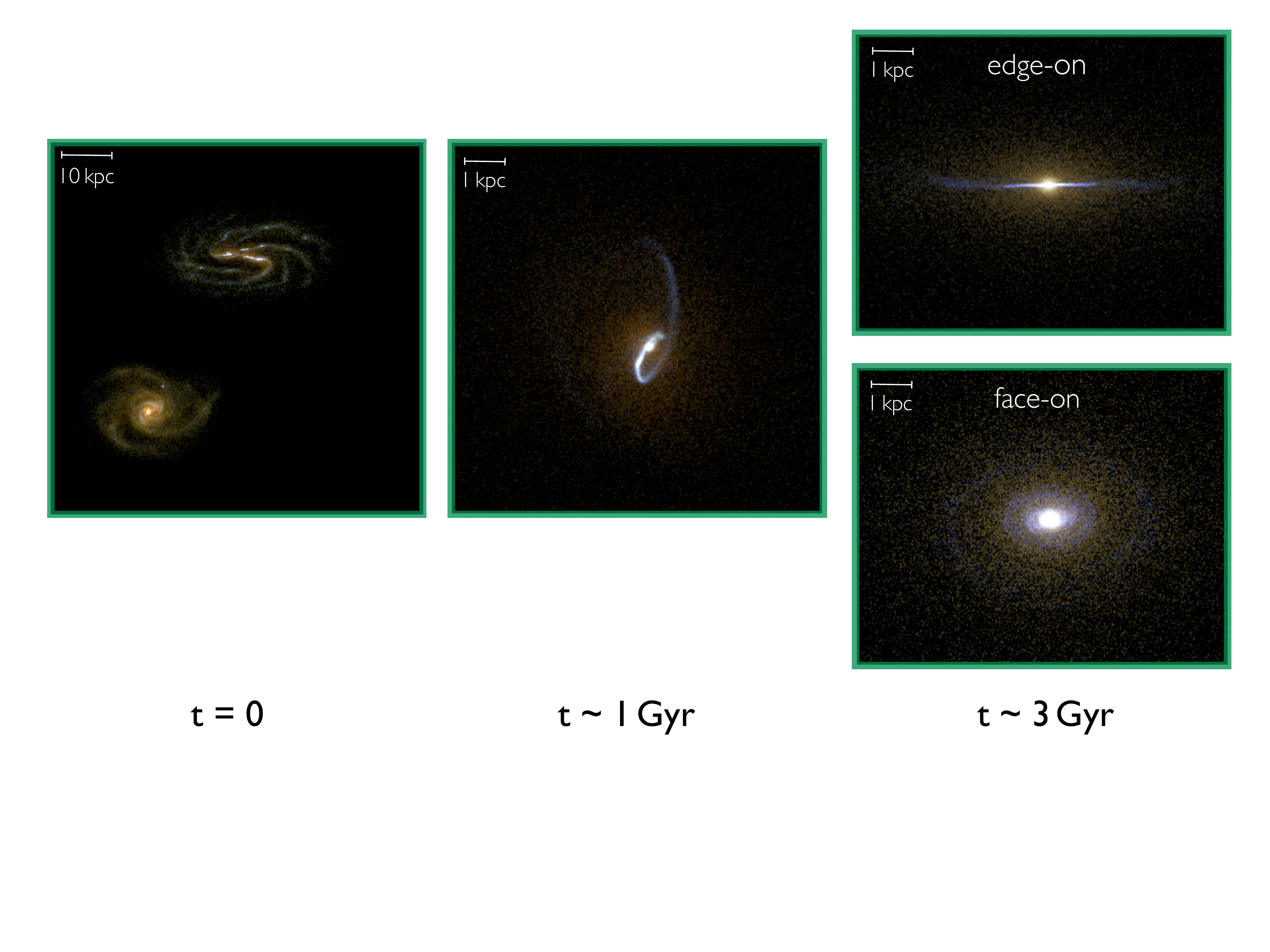}
\caption{Three stages of the major merger between two giant spiral galaxies, Sa and Sd, simulated by GalMer (inclination=75$^{\circ}$, pericentre=16\,kpc, v$_0$=200\,km/s, prograde encounter). Shown are RGB synthetic images combining simulated \textit{K}-band (red), SDSS \textit{g}-band (green), and GALEX near-UV (blue).}
\label{fig:sequence}
\end{center}
\end{figure*}

We select the major mergers (mass ratios 1:1 to 1:3) from GalMer involving all possible combinations of two spiral progenitors (Sa, Sb, Sd) that result in a realistic and dynamically relaxed remnant.
We have imposed some qualitative criteria on structural, kinematic, SFR, and gas content levels (to match typical observed values of S0s; Eliche-Moral et al.~in preparation), and we have performed a visual morphological classification to identify which remnants would have been classified as S0-like by observers (i.e.~as disc galaxies without noticeable spiral arms).
To do that, we have created mock photometric images of the resulting galaxies in several bands (\textit{B}, \textit{V}, \textit{R}, \textit{I}, and \textit{K}), using a mass-to-light ratio which considers the stellar mass, age, and metallicity of each simulation particle according to \citet{2003MNRAS.344.1000B}, with a Chabrier IMF and the Padova evolutionary tracks. This provides us with a final sample of 67 S0-like remnants, on which we have performed bulge-disc-(bar) photometric decompositions analogous to observers.

\subsection{Results on photometric scaling relations}
\label{photodiscussion}

\begin{figure*}[t]
\begin{center}
\includegraphics[trim=30 40 30 120,clip,height=0.48\textwidth]{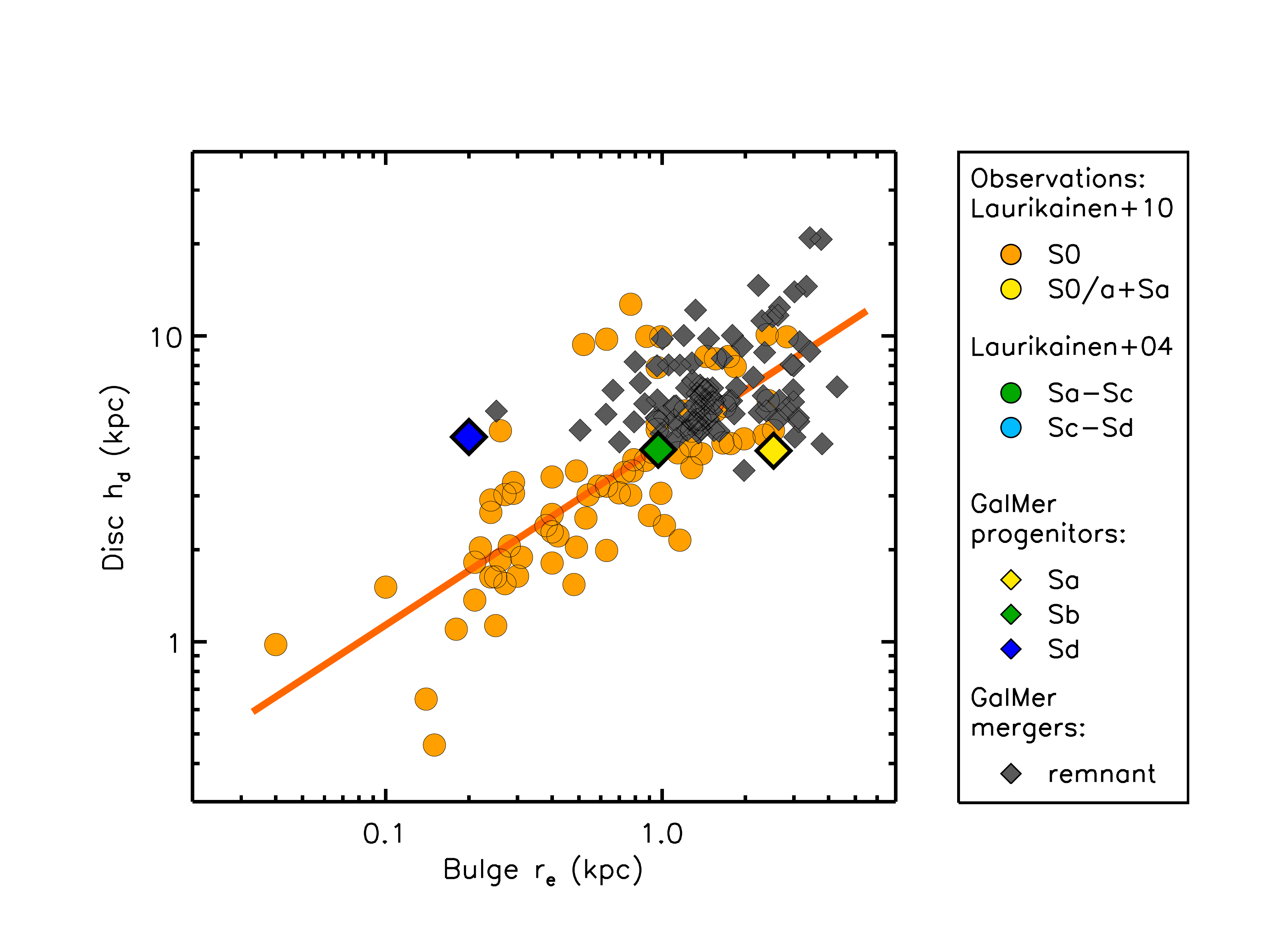}
\caption{Highlight from Fig.\,5 in \citet{2015A&A...573A..78Q}, making it clear how the major merger remnants that we have studied (black diamonds) overlap the observations of S0s (orange) from \citet{2010MNRAS.405.1089L}; observational spirals are not shown here for clarity purposes, but they display a larger scatter than lenticulars.
}
\label{fig:photom}
\end{center}
\end{figure*}

Except for pathological cases (such as some coplanar orbits), discs in the major mergers leading to S0s are first destroyed but quickly rebuilt during the last stages of the merging sequence. This is visually illustrated by Fig.\,\ref{fig:sequence} (a major merger of two giant spirals, Sa and Sd), and  demonstrated in a more qualitative way in Fig.\,7 from \citet{2015A&A...573A..78Q}. The original structure is mostly destroyed in the intermediate stages (\textit{t}$\sim$1\,Gyr), leading to spectacular tidal tails, but, in the end (\textit{t}$\sim$3\,Gyr), the whole system ends up relaxing into a passive disc with a small but clear bulge: a lenticular galaxy.

Our photometric decompositions of these major merger remnants allow us to populate the same diagrams studied by observers. Fig.\,\ref{fig:photom} shows the relationship between scalelength of the disc and the effective radius of the bulge, which is one of the 9 photometric planes studied in \citet{2015A&A...573A..78Q}. It demonstrates how, contrary to expectations, not only can discs survive major mergers, but also they can lead to lenticular galaxies in which disc and bulge sizes scale relative to each other, soon ($\sim$1\,Gyr) after the merging process is completed. In other words, the observational scaling relation tells us that, for a given bulge size, only a certain range of disc sizes are allowed by nature in real lenticulars, and what we have found is that the S0s emerging from our major merger simulations have discs and bulges which are also coupled in that way, in agreement with photometric observations. The fact that the simulated major merger remnants only cover the top-right end of the plot on Fig.\,\ref{fig:photom} is not surprising, because the original spiral galaxies from the GalMer simulations are already very massive, and, by bringing two of them together, the total stellar mass (and, indirectly, the associated disc and bulge sizes) are expected to grow.

In \citet{2015A&A...573A..78Q} we show how the S0-like remnants from the major mergers that we have studied obey similarly tight relations in other photometric planes (involving bulge-to-total luminosity ratio, bulge effective radius, disc scalelength, total bulge and disc magnitude, and S\'ersic index). This confirms the compatibility of the merger remnants with photometric observations of S0s; the reader is referred to the paper for those additional planes and more thorough discussion.

\section{Creating lenticulars with major mergers: kinematics}
\label{lambda}

\begin{figure*}[t]
\begin{center}
\includegraphics[trim=30 40 30 120,clip,height=0.48\textwidth]{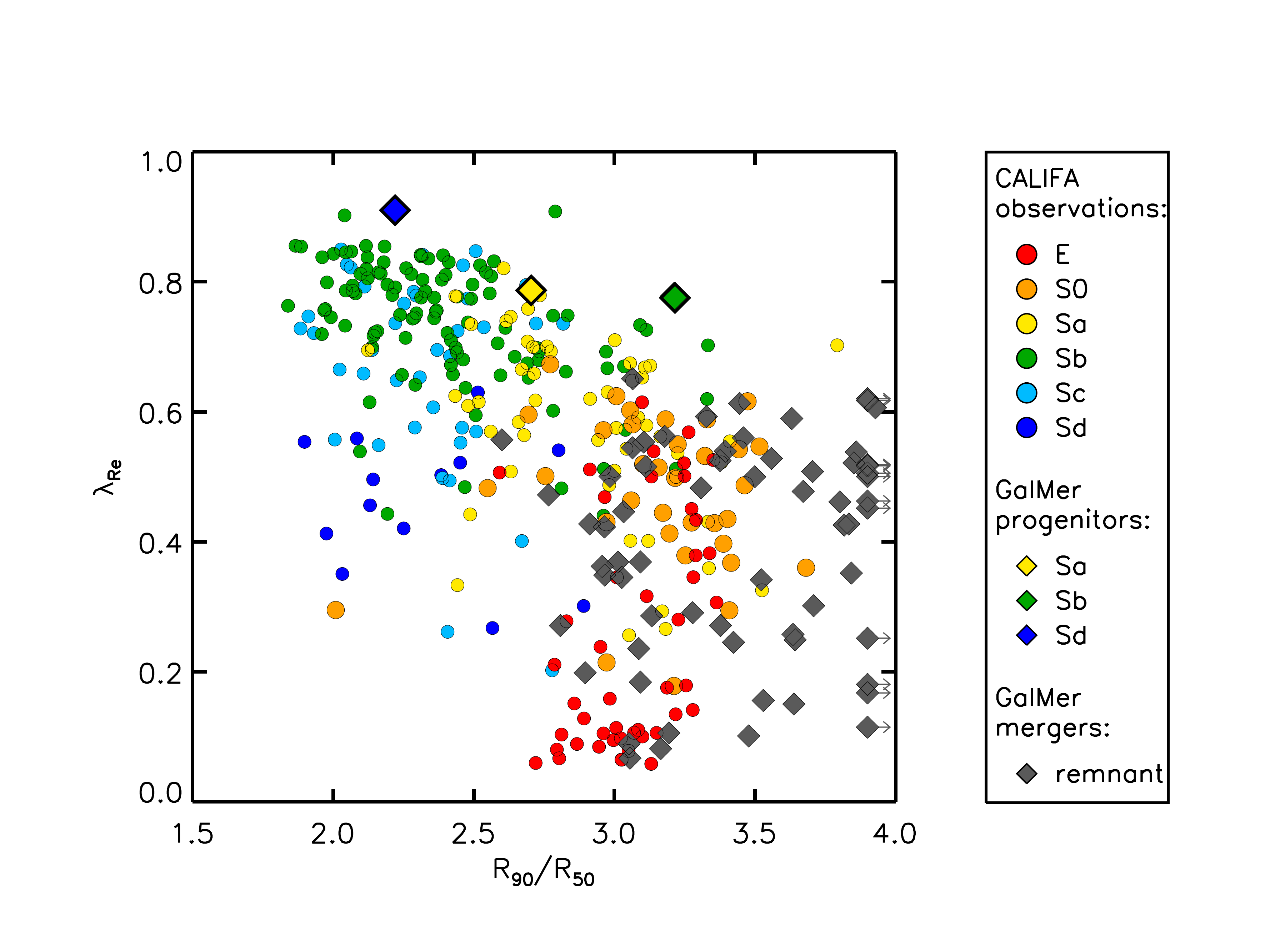} \caption{Stellar angular momentum ($\lambda_\mathrm{Re}$) plotted against concentration (\textit{R}$_{90}$/\textit{R}$_{50}$) for the GalMer simulations (original models and merger remnants), in comparison to CALIFA galaxies. All the parameters correspond to an edge-on view.}
\label{fig:kinem}
\end{center}
\end{figure*}

The CALIFA team has recently presented a new diagnostic diagram, the $\lambda_\mathrm{Re}$--concentration plane, which constitutes a strong objection to the idea that S0s are mostly faded spirals (\citep{2015IAUS..311...78F}, van de Ven et al. in preparation). Late-type spirals as a population (Sb, Sc, Sd) are clearly incompatible with S0s when both angular momentum ($\lambda_\mathrm{Re}$) and concentration (\textit{R}$_{90}$/\textit{R}$_{50}$) are simultaneously taken into account.  Simple fading is not expected to significantly change the angular momentum of the galaxy, which is why the CALIFA findings contradict the idea that most S0s are spirals which have had their gas removed.
 
Fig.\,\ref{fig:kinem} compares the distribution of our remnant S0s with that of real galaxies from CALIFA in 
$\lambda_\mathrm{Re}$ versus \textit{R}$_{90}$/\textit{R}$_{50}$,
showing that S0s resulting from the simulated major mergers are consistent with real lenticulars \citep{2015A&A...579L...2Q}. We calculate $\lambda_e$ according to \citet{2011MNRAS.414..888E}, out to the effective radius of the galaxy (\textit{R}$\leq$\textit{R}$_\mathrm{e}$). 
The light concentration is estimated via the Petrosian ratio \textit{R}$_{90}$/\textit{R}$_{50}$, defined as the ratio of the radii enclosing 90\% and 50\% of the Petrosian flux, measured on the 1D azimuthally averaged profile of our mock SDSS \textit{r} band images.  We assume that the simulated merger remnants are observed at the median distance of the 300 CALIFA galaxies in Falc{\'o}n-Barroso et al. (in preparation), \textit{D}$_{\mathrm{median}}$ = 67\,Mpc, and we simulate the same spatial resolution and field of view as that survey.

In the $\lambda_\mathrm{Re}$--concentration plane (Fig.\,\ref{fig:kinem}), Sb and Sc galaxies observed by CALIFA accumulate towards the top-left corner; Sa galaxies are more scattered, extending towards the area populated by S0 and E galaxies in the bottom-right. This first CALIFA sample has a very limited number of Sd galaxies, but their overlap with S0s is also minimal.
From this plot, we conclude that major mergers can make spiral progenitors evolve from the region of high $\lambda_\mathrm{Re}$--low concentration into realistic S0 galaxies of lower $\lambda_\mathrm{Re}$ and higher concentration, following a trend analogous to that observed by CALIFA.
Even though many Sa galaxies are compatible with S0s in this plane, the change is very significant from later-type spirals (Sb, Sc, Sd) into S0s, and this is where major mergers are required (or any other mechanisms capable of changing both the angular momentum and the concentration).

We briefly summarise some caveats next. It is true that a significant number of our S0 merger remnants are outliers in terms of concentration. However, we understand that this is because we are looking at the remnants in the simulations \textit{short after coalescence} (due to the total simulated time of $\sim$4\,Gyr, so typically $\sim$1\,Gyr after full merger).
Provided that we consider a mass-to-light ratio which takes the age and metallicity of the stellar particles in the simulation into account, a central starburst (which is very often associated with mergers, see \citep{1991ApJ...370L..65B}) will lead to a very high light concentration due to young stars even $\sim$1\,Gyr after the merger has finished, which reflects on the \textit{R}$_{90}$/\textit{R}$_{50}$ parameter. In this context, it is not surprising that the most extreme concentration outliers (\textit{R}$_{90}$/\textit{R}$_{50}$\,$>$ 4.5) correspond to the encounters involving the highest initial gas fractions ($>$ 20\%), and the highest fractions of merger-triggered star formation (new stars $>$ 15\% of total stellar mass). Our paper emphasises the importance of choosing an appropriate mass-to-light ratio when translating simulations of galaxy mergers into the observational parameter space, especially when it comes to calculating concentration parameters.
Finally, the fact that the progenitor Sa is less concentrated than the Sb galaxy is not necessarily unrealistic (there is a considerable number of observational Sa datapoints with lower \textit{R}$_{90}$/\textit{R}$_{50}$ than a many Sb galaxies); bulge-to-total ratio and light concentration are often invoked as a proxy for morphology, but there is not a one-to-one match from one to the other (see e.g.~\citep{2008MNRAS.388.1708G}).


\section{Concluding remarks}

Some simulators had already suggested that major mergers do not always necessarily destroy discs (\citep{2002MNRAS.333..481B, 2005ApJ...622L...9S, 2004ApJ...606...32R, 2006ApJ...645..986R, 2009ApJ...691.1168H, 2011MNRAS.415.3750M}; but see the doubts raised by \citet{2011ApJ...730....4B} about the need to include supersonic turbulence in such simulations of gas-rich mergers). Here, we have explored the possibility that major mergers could lead to the creation of lenticular galaxies using \textit{N}-body numerical simulations of dissipative binary galaxy mergers from the GalMer database.

In \citet{2015A&A...573A..78Q} we proved that not only can discs survive major mergers, but also they can  produce remnants with bulge-disc coupling in perfect agreement with observations of S0s \citep[e.g.][]{2010MNRAS.405.1089L}. One of those planes (disc scaleheight versus bulge effective radius) has been presented in Fig.\,\ref{fig:photom} and discussed in Sect.\,\ref{photodiscussion}, proving that the S0-like remnants from the major merger simulations have discs and bulges that obey analogous scaling relations to those observed in real lenticulars.

In \citet{2015A&A...579L...2Q} we showed that, in addition, the systematic offset between spirals and S0s in the angular momentum ($\lambda_\mathrm{Re}$)--concentration (\textit{R}$_{90}$/\textit{R}$_{50}$) plane reported by the CALIFA team can be explained by major mergers. The main figure from the paper has been reproduced here (Fig.\,\ref{fig:kinem}) and briefly commented in Sect.\,\ref{lambda}.
 Additionally, using the same numerical simulations, we have recently found that those S0-like remnants have discs with antitruncations in agreement with those observed in real S0s (matching their strong scaling relations, see \citet{2014A&A...570A.103B}).
 If we combine all these photometric and kinematic results, major mergers appear as a viable mechanism to transform spirals into lenticulars.

Naturally, this does not mean that major mergers are responsible for the emergence of \textit{all} lenticular galaxies in the local Universe, and the relative contribution of (major) mergers is probably strongly dependent on environment. 
In the near future, we plan to extend this investigation to mergers of other mass ratios, and further explore the synergies between simulations of galaxy mergers and the 3D observations from CALIFA.


\acknowledgments{Acknowledgments}

The authors would like to acknowledge the GalMer team for creating such a powerful tool, with especial thanks to Paola Di Matteo for her kind support with the database. MQ, GvdV and JFB acknowledge financial support to the DAGAL network from the People Programme (Marie Curie Actions) of the European Union's Seventh Framework Programme FP7/2007- 2013/ under REA grant agreement number PITN-GA-2011-289313. CEM acknowledges support from the Spanish Ministry of Economy and Competitiveness (MINECO) under projects AYA2012-31277 and AYA2013-48226-C3-1-P. JMA acknowledges support from the European Research Council Starting Grant (SEDmorph; P.I. V. Wild).


\authorcontributions{Author Contributions}

The two papers that these conference proceedings are based on \citep{2015A&A...573A..78Q,2015A&A...579L...2Q} have been lead and written by the first author, Miguel Querejeta, while all co-authors listed here have substantially contributed to the reported work.


\conflictofinterests{Conflicts of Interest}

The authors declare no conflict of interest.

\newpage


\bibliography{S0s.bib}{}
\bibliographystyle{mdpi}


%


%

\end{document}